\documentclass[useAMS,usenatbib,usegraphicx]{mn2e}
\usepackage{xspace}
\usepackage{amssymb}

\newcommand{\SM}{\textsl{SM}\xspace}
\newcommand{\NM}{\textsl{NM}\xspace}
\newcommand{\Nbody}{\textsl{N}-body\xspace}

\title[Chaotic mixing and the evolution of triaxial galaxies]
{Chaotic mixing and the secular evolution of triaxial cuspy galaxy models built with Schwarzschild's method}
\author[E.~Vasiliev, E.~Athanassoula]
{E. Vasiliev$^{1,2,3}$\thanks{E-mail: eugvas@lpi.ru (EV), lia@oamp.fr (EA)}, E. Athanassoula$^{3}$\\
$^{1}$Lebedev Physical Institute, Leninsky prospekt 53, Moscow, Russia\\
$^{2}$Rochester Institute of Technology, 76 Lomb Memorial drive, Rochester, NY 14623, USA\\
$^{3}$Laboratoire d'Astrophysique de Marseille (LAM), UMR6110, CNRS/Universit\'e de Provence, 
38 rue Joliot Curie, 13388 \\ Marseille C\'edex 13, France}

\begin{document}

\date{August 8, 2011}

\pagerange{\pageref{firstpage}--\pageref{lastpage}} \pubyear{2011}

\maketitle

\label{firstpage}

\begin{abstract}
We use both \Nbody simulations and integration in fixed potentials to
explore the stability and the long-term secular evolution of
self-consistent, equilibrium, non-rotating, triaxial spheroidal
galactic models. More specifically, we consider Dehnen models  built
with the Schwarzschild method.
We show that short-term stability depends on the degree of velocity anisotropy 
(radially anisotropic models are subject to rapid development of radial-orbit instability). 
Long-term stability, on the other hand, depends mainly on the
properties of the potential, and
in particular, on whether it admits a substantial fraction of strongly chaotic orbits. 
We show that in the case of a weak density cusp ($\gamma=1$ Dehnen model) the \Nbody model is 
remarkably stable, while the strong-cusp ($\gamma=2$) model exhibits substantial evolution 
of shape away from triaxiality, which we attribute to the effect of chaotic diffusion of orbits.
The different behaviour of these two cases originates from the different phase space structure of 
the potential; in the weak-cusp case there exist numerous resonant orbit families that impede 
chaotic diffusion.
We also find that it is hardly possible to affect the rate of this evolution by altering the 
fraction of chaotic orbits in the Schwarzschild model, which is explained by the fact 
that the chaotic properties of an orbit are not preserved by the \Nbody evolution. 
There are, however, parameters in Schwarzschild modelling that do
affect the stability of an \Nbody model, 
so we discuss the recipes how to build a `good' Schwarzschild model.
\end{abstract}


\section{Introduction}  \label{sec_intro}

The Schwarzschild method is an important tool for constructing self-consistent 
equilibrium models of galaxies, when the system does not have
an analytic distribution function \citep{Schw79}. 
Models of triaxial galaxies are of particular interest, since they often admit 
a large fraction of chaotic orbits.

The aim of the Schwarzschild method is to construct self-consistent equilibrium models, 
in which orbits in a given potential are arranged so that the resulting density 
matches the potential via the Poisson equation. 
On the other hand, dynamical stability of these models is out of the
scope of the Schwarzschild method. 

There are two reasons why such a model might turn out to be non-stationary. 
First this could be due to dynamical instabilities, such as the radial-orbit instability 
\citep[e.g.][]{PolyachenkoShukhman81},
which manifest themselves on a rather short time, typically of the
order of a crossing time.
The other reason concerns more gradual evolution and is related to the existence of chaotic orbits in 
most non-integrable potentials, especially those having a high central mass concentration 
(density cusp or black hole). 
Orbits in the Schwarzschild model (\SM) are regarded as stationary `building blocks', 
which ensures that the distribution function of the whole model satisfies the 
time-independent collisionless Boltzmann equation \citep{BT}, $df/dt=0$.
By construction, orbits are evolved for a certain time, typically of order 
$10^2$ dynamical times, which is expected to be sufficient to sample the 
available phase space. This is a reasonable assumption for regular orbits, 
which fill their invariant tori more or less uniformly during this time.
But chaotic orbits have a much larger region of available phase space (of higher dimensionality) 
and may sample it in a very non-uniform way, remaining in a confined portion of this region for 
many hundreds of dynamical times. 
This may lead to the effect of chaotic diffusion, when these orbits eventually escape to 
a different region of phase space and, accordingly, change their shape in configuration space, 
which leads to the breakdown of the self-consistency of the model. 

The stability of models built with Schwarzschild's method has been tested by \Nbody simulations 
as early as in \citet{SmithMiller82}, albeit with a rather rough \Nbody code and on a short timescale.
\citet{Zhao96} confirmed the stability of a galactic bar model using the self-consistent field (SCF) method.
More recently, \citet{Antonini09} showed that the radial-orbit
instability (ROI) exists also in triaxial 
systems, if the velocity anisotropy coefficient $\beta \gtrsim 0.3$. 
\citet{Wu09} explored the existence and stability of triaxial Dehnen models in MOND gravity and found 
it to be acceptable. Nevertheless, their \Nbody models initially displayed some evolution towards a more 
equilibrium configuration, presumably due to difficulties associated with constructing 
self-consistent models in MOND gravity.

On the other hand, the effect of chaotic diffusion on the evolution of
a \SM is usually assumed to 
lead towards a more spherical mass distribution.
Support for this conjecture comes from the fact that the `average' shape of a fully chaotic orbit 
(which fills almost ergodically the equipotential surface, excluding the parts of phase space 
occupied by regular orbits) is generally rounder than the equidensity surface.
\citet{Schw93} confirmed this effect for a scale-free logarithmic potential, corresponding to 
a $\gamma=2$ density cusp, by creating a self-consistent \SM and then following the ensemble of 
orbits for an interval of time three times longer than was used in creating the model.
He recorded the shape obtained from the superposition of these orbits
for this longer interval and  
concluded that it was evolving towards sphericity, but that this change was quite small. 

A number of studies used \Nbody simulations to address the long-term stability of triaxial models 
created by various methods, and their evolution caused by chaotic orbits.
\citet{Holley01} constructed an equilibrium triaxial model with axial ratio $a:b:c=1:0.85:0.7$ 
based on the spherical Hernquist model ($\gamma=1$) by applying artificial squeezing along two axes,
and evolved it with a SCF \Nbody code for $\sim 8$ half-mass 
dynamical times to confirm that there is no significant change in shape.

Later, \citet{Holley02} extended their study to include a supermassive black hole with mass 
$M_\bullet$ equal to 0.01 of the total model mass. The growth of this central point mass destabilizes the 
population of box orbits and converts most of them into rather strongly chaotic ones, which, 
in turn, quickly drives the inner regions of the model towards almost
spherical shape. Again using SCF \Nbody code, 
\citet{Kalapotharakos04} explored the dependence of such evolution on
the black hole mass, 
and found that in all cases there existed a large fraction of chaotic orbits, but the overall shape 
of the model substantially evolved only for $M_\bullet>0.005$, when these orbits were more strongly
chaotic (as measured by Lyapunov exponents), while smaller $M_\bullet$ did not cause much 
secular evolution.
\citet{Muzzio09} created a strongly triaxial model with a $\gamma\simeq 1$ cusp by cold collapse 
and confirmed its stability by quadrupolar (a restricted variant of SCF) \Nbody code over several 
hundred crossing times, despite having large ($\gtrsim 50\%$) fraction of chaotic orbits.
All these results are based on \Nbody modelling, and the methods used
for creating their initial conditions cannot make a system with predefined 
properties, unlike the iterative method of \citet{Rodionov09}, discussed in 
Section~\ref{sec_compare_iterative}.

\citet{PoonMerritt04} constructed models for triaxial scale-free cusps 
with $\gamma=1$ and $\gamma=2$ around a black hole using the Schwarzschild method. 
Their solutions contained about half of the mass in chaotic orbits, but nevertheless were found 
to be reasonably stable when evolved by an \Nbody tree-code during $6\,T_{dyn}$ 
(measured at few times the black hole influence radius), except for prolate ($T=0.75$) models.
They conclude that their chaotic orbits were sufficiently mixed during the $T=100$ orbital times 
used for integration in \SM, so that they represented reasonably stationary building blocks. 
However, the time interval was quite short (only a few dynamical times for orbits outside the 
radius of influence, where most chaotic orbits are found), and the change of axial ratios was 
small yet non-negligible. In addition, they followed the evolution of models in a fixed smooth 
potential for $\sim 100$ dynamical times, and found no change in
shape, as is reasonable to 
expect, given that orbits in the \SM were evolved for a similar
time. Another reason for the less apparent 
shape evolution in their models is that for scale-free potentials the equipotential surface is not 
much rounder than the equidensity surface, so even `fully chaotic' orbits do support the necessary 
model shape to some degree.

A somewhat different issue is addressed in \citet{Valluri10}.
They investigated the change of shape of triaxial dark matter halos in response to the growth of a
compact mass in the centre, that is, due to an adiabatic change of the potential. 
They compared the orbit population of the models 
before the growth of the central mass, after the growth (intermediate stage), and after an adiabatic 
`evaporation' of this mass (final stage).
They used an \Nbody tree-code both to follow the evolution of the `live' system and to analyze the 
properties of the orbits in the `frozen' \Nbody potential. 
\citet{Valluri10} found that, in the case with no strong central mass concentration, 
the evolution of orbit shapes is mostly reversible, despite the fact that many of the orbits 
are chaotic in the intermediate stage, and attributed this reversibility to resonant trapping 
of orbits in the course of the slow change of the potential. 

We continue and extend these studies in two interconnected aspects. 
Namely, we perform \Nbody simulations of triaxial cuspy Dehnen models built with 
the Schwarzschild method, and find that, in the cases when there is no rapid onset of 
radial-orbit instability, they are remarkably stable over many dynamical times. 
There is, however, a long-term evolution of the shape of these models
and we find that it is caused by the
influence of chaotic orbits. 

We begin by introducing our triaxial Dehnen models constructed with the Schwarzschild method, 
and briefly review their properties in section~\ref{sec_models_intro}.
Then in section~\ref{sec_chaos_general} we review the concept of chaos,
in particular, the distinction between regular, weakly (sticky) chaotic and strongly chaotic orbits, 
and the quantities that are intended to represent chaotic properties
of an individual orbit.

The long-term evolution of orbits in a fixed potential is the subject of 
section~\ref{sec_chaotic_shape_change}. It turns out that in the potentials considered here, most 
chaotic orbits are in fact quite sticky, and the timescale for chaotic diffusion and the associated 
change of orbit shapes is rather long, of order $10^3$ dynamical times. 
However, in the strong-cusp model and in the outer parts of the weak-cusp model this stickiness is 
less prominent, and the chaotic orbits do exhibit evolution towards a more spherical shape.
We also discuss how our findings can be explained by the amount of `complexity' 
of the phase space, that is, the presence and importance of resonant orbit families. 

Next, we describe our \Nbody experiments which basically confirm the expectations derived from 
the fixed-potential evolution (section~\ref{sec_nbody_evolution}). 
Unless the velocity anisotropy in the model is sufficiently biased towards radial velocities to 
allow for a rapid development of the radial-orbit instability, the 
shape of the density distribution in the
\Nbody model (\NM) remains in agreement with that of the original Schwarzschild model (\SM) for 
a long time. The weak-cusp model is essentially stable over the course
of the simulation, while for
the strong-cusp case there is a gradual evolution towards a more spherical (or, rather, oblate 
axisymmetrical) shape, which we attribute to chaotic diffusion.

In addition, a very important, and somewhat disappointing, finding is that we are hardly able to 
control the degree of evolution due to this chaotic diffusion by
altering the properties of the \SM.
This is because the orbits in the \NM, while basically resembling
their counterparts in the \SM in  
shape and orbital class, do not inherit the attribute of chaoticity
from the smooth-potential model. 
This argues that the overall evolution of the \NM is determined mostly by the gross properties of the potential, 
rather than by any specific arrangement of orbits (as long as this satisfies self-consistency).
Nevertheless, we do present recipes for building better \SM (in the sense that the corresponding \NM 
are more stable) in section~\ref{sec_schw_variants}.

In section~\ref{sec_compare_iterative} we compare the evolution of models built with the 
Schwarzschild method with that of a model constructed with the iterative method \citep{Rodionov09}, 
which is completely different from the \SM as it relies on the `guided
evolution' of an \Nbody model 
towards a specific equilibrium. 
We show that, despite conceptually being very different, they perform similarly in terms of 
stability of the resulting model, and have similar properties of
ensemble of orbits.  

Finally, we present our conclusions.

\section{Schwarzschild models}  \label{sec_models_intro}

In this paper we restrict our attention to non-rotating three-dimensional systems with mild flattening.
Namely, we consider two variants of the triaxial Dehnen model, with density profile
\begin{equation}  \label{eq_modelrho}
\rho(r) = \frac{(3-\gamma)M}{4\pi a b c} \frac{1}{m^{\gamma}(1+m)^{4-\gamma}},
\end{equation}
where $m=[(x/a)^2+(y/b)^2+(z/c)^2]^{1/2}$ is the elliptic radius. We adopt dimensionless units in which 
$M=1$, $a=1$, $G=1$ (which also fixes the time unit)%
\footnote{Note that for dimensionless radius we use the long-axis scale 
radius $a$ and not the half-mass radius, the latter being $2.4a$ for $\gamma=1$ 
and $a$ for $\gamma=2$ models, measured along the $x$ axis.}, 
and choose the axial ratios $b/a=\sqrt{5/8}\approx 0.79$, $c/a=1/2$, which corresponds to a
triaxiality parameter $T=(a^2-b^2)/(a^2-c^2)$ equal to $1/2$ (maximal triaxiality)%
\footnote{$T=0$ corresponds to oblate and $T=1$ to prolate axisymmetric models, so $T=1/2$ 
is called `maximally triaxial'.}.
For the cusp slope $\gamma$ we choose two values: $\gamma=1$ (weak cusp) and $\gamma=2$ (strong cusp). 
It turns out that there are substantial differences between the two models, which will be discussed in 
the following sections.

All integration times and orbit frequencies in the \SM are measured in
units of dynamical time 
$T_{dyn}(E)$, which is defined as the period of the long-axis orbit
with the given energy:
$T_{dyn} = 4\int_0^{r_{max}(E)} [2(E-\Phi(r))]^{-1/2}\, dr$.
Other papers often use the period of the $x-y$ plane closed loop orbit, which is somewhat shorter, 
but our definition has the advantage that all natural frequencies of orbits are greater than 
$T_{dyn}^{-1}$. For our models $T_{dyn}(r)$ may be approximated as $4.4(r^{1/2}+r^{3/2})$ 
for the weak-cusp case and $4.4(r^2+r^3)^{1/2}$ for the strong-cusp
case, where time and length units are dimensionless
time units.

The models were constructed using 50 radial shells, 2400 grid cells and $3\cdot10^4$ orbits 
evolved for 500 dynamical times. 
The initial conditions for the orbits were assigned randomly using the following recipe:
sample the elliptical radius uniformly in the value of enclosing mass, 
randomly choose angles to put the orbit onto the equidensity ellipsoid, 
and assign each component of the velocity according to a Gaussian distribution 
with the dispersion of the equivalent spherical model. 
This is different from the traditionally used method of sampling initial conditions 
at a small number of energy levels and in fixed positions on the grid; we find that 
a random position and, more importantly, continuous distribution in energy yields better models.

In the following three sections we will discuss mainly one variant for each model, namely the 
`unconstrained' variant with no restrictions on the orbit population
or on the fraction of chaotic orbits,
and with a velocity anisotropy $\beta = 1-\frac{\sigma_t^2}{2\sigma_r^2}$ \citep{BT}
varying from 0 in the centre to $\sim 0.6$ in the outer parts. 
We consider the effect of variation of these parameters in section~\ref{sec_schw_variants}.

\section{General remarks on the chaotic properties of an orbit}  \label{sec_chaos_general}

There are several methods for quantifying the degree of chaoticity of a given orbit, which may be 
classified into two broad groups. One group deals with a single orbit and considers its spectrum, 
that is, the Fourier transform of some quantity along the trajectory
(e.g. the $x$ coordinate, or the distance from the centre) sampled at
equally spaced moments of time. 
It is based on the fact that all regular orbits are multiply-periodic, with their spectra 
containing in 3D only linear combinations of no more than three fundamental frequencies, 
these frequencies being constants of the motion in a time-independent potential.
Any deviation from this simple spectrum is an indication of chaos. 
A quantitative estimate of chaos may be obtained either as the number of spectral 
lines containing a specified fraction (say, 0.9) of the total power \citep{KandrupEB97}, 
or as the variation in these fundamental frequencies calculated e.g. from the first and 
second halves of the integration time \citep{Laskar93, ValluriMerritt98}, 
dubbed `frequency diffusion rate' (FDR).

The second group of methods considers the deviation of nearby orbits; 
the orbit in question is accompanied by one or more adjacent orbits, 
and the evolution of deviation vectors is tracked.
In the case of a regular orbit these deviation vectors should grow no faster than linearly, 
and if more than one such vector is considered, they should remain correlated in direction.
Conversely, in the chaotic case the deviation starts to grow exponentially after some time. 
This class includes methods based on the calculation of Lyapunov exponents $\Lambda$ 
and alignment indexes \citep[e.g.][]{Skokos10}.
There is quite a good correspondence between 
FDR and $\Lambda$ as chaos indicators in a smooth potential. 
To construct \SM with different fraction of chaotic orbits (Section~\ref{sec_schw_variants}), 
we use here the Lyapunov exponents to distinguish between regular and chaotic orbits.

For an \Nbody system, however, all trajectories have large positive
Lyapunov exponents.  
Their timescale for exponential deviation is a fraction of the crossing time, 
and furthermore, they do not decrease with increasing $N$ 
\citep{KandrupSideris01, SiderisKandrup02, ValluriMerritt00}%
\footnote{A method to estimate \textit{true} Lyapunov exponent for 
an orbit in a frozen-\Nbody system has been proposed by \cite{KandrupSideris03}, but it is unclear whether 
it can be easily applied to live simulations.}.
This does not preclude a more regular behaviour of the system with more particles, 
since Lyapunov exponents, by definition, measure the growth rate of infinitely small 
perturbations, but do not tell by what distance two nearby orbits will be separated after a finite time. 
It appears that if the phase space has a complex structure of stable resonant islands, 
then chaotic orbits usually tend to be confined to small regions of the entire phase space, 
demonstrating the so-called phenomenon of stickiness, and resemble regular orbits for many 
dynamical times \citep[e.g.][]{Contopoulos71, Contopoulos02, ValluriMerritt00, HarsoulaKalapotharakos09}.  

Therefore, for the purpose of comparing the chaotic properties of orbits evolved in a smooth 
potential and in an \Nbody model, we will restrict ourselves to the first class of 
chaos detection methods. 
Namely, we use the following definition for the frequency diffusion rate (FDR):
\begin{equation}  \label{eq_fdr}
\Delta \omega \equiv \frac{1}{3} \sum_{i=1}^3 
  \frac{|\omega_{i}^{(1)}-\omega_{i}^{(2)}|}{(\omega_{i}^{(1)}+\omega_{i}^{(2)})/2}
\end{equation}
Here $\omega_i^{(1)}$ and $\omega_i^{(2)}$ are the leading frequencies in Cartesian coordinates 
($i=x,y,z$) for the first and the second halves of the integration
time, respectively%
\footnote{This is different from what was used in \cite{ValluriMerritt98}, 
where only the largest of the three differences $\Delta\omega_i$ was tracked. 
We find that their definition may sometimes exhibit unwanted
fluctuations, when the spectrum has two distinct lines of comparable 
amplitudes: one may become the leading frequency for the first half, the other 
for the second half, and the difference will be large. 
(Of course, this still means that the spectrum experiences changes, but not necessarily that rapidly). 
So we use the averaged value for all three coordinates, and furthermore, discard the lines 
with relative variation greater than 0.5 (which are indeed rare).}.
We adopt $\Delta\omega=10^{-3}$ as the threshold separating regular from chaotic orbits, 
which roughly corresponds to the distinction based on the Lyapunov exponent, 
if both are measured on the interval of $100\,T_{dyn}$.

Unfortunately, FDR itself is not a strictly defined quantity: if we measure $\Delta\omega$ for two 
successive integration intervals, or even for a somewhat different duration of time (say, 100 and 
110 $T_{dyn}$), or change the sampling rate, this quantity may change by a factor of few%
\footnote{However, if the sampling rate is changed by an integer factor, the frequencies are likely 
to remain the same, as tested in \cite{Valluri10}.}.
This is not surprising, since this quantity by definition measures the difference in two 
'instantaneous' values of a fluctuating variable $\omega$, and is itself a random quantity 
with an uncertainty of about $0.3-0.5$ orders of magnitude.

The shape of an orbit in configuration space may be described by its inertia tensor components, 
\begin{equation}  \label{eq_inertia}
I_{ij} \equiv \frac{1}{N_s}\sum_{n=1}^{N_s} x_i^{(n)} x_j^{(n)} \,,
\end{equation}
where $x_i^{(n)}$ is the $i$-th component of the $n$-th sampling point of the trajectory. 
If we consider only diagonal components, 
$\sqrt{I_{ii}}$ gives an estimate of the extent of this orbit in the $i$-th coordinate.
Likewise, the quantities 
\begin{equation}  \label{eq_flattening}
S_i \equiv \frac{I_{ii}}{I_{xx}+I_{yy}+I_{zz}} \;,\quad i=x,y,z
\end{equation}
describe the orbit shape (flattening in the $i$-th direction). 
Tracing their change over time may be used to estimate the shape evolution.

Finally, it is necessary to note that the accuracy of determination of orbital frequencies 
is limited by the accuracy of energy conservation (typically $\Delta\omega \simeq |\Delta E/E|$).
While this is not a limitation in the case of a smooth time-independent potential 
(the error in energy conservation in the integrator is $\lesssim 10^{-9}$), it becomes 
a major obstacle when we come to analyzing the orbits in a live \Nbody simulation, 
where particles experience random variations in energy, e.g. due to two-body relaxation.

\section{Secular shape evolution induced by chaotic orbits}  \label{sec_chaotic_shape_change}

The construction of equilibrium models by the Schwarzschild method relies on finding time-independent 
`building blocks' as required by Jeans' theorem. Regular orbits obviously satisfy this requirement 
(as long as we ensure sufficiently uniform coverage of the invariant torus), 
but for chaotic orbits it's not the case.
It has been suggested that all chaotic orbits of a given energy may be in fact considered 
as representatives of one `super-orbit' (in 3D, the chaotic part of the phase space at a given 
energy is one interconnected region, the so-called Arnold web), 
and hence, if we average all these orbits, the resulting building block may also be used 
as a time-independent unit. 
In practice, however, the usefulness of this approach is doubtful, for a number of reasons.
First of all, it is not easy to distinguish unambiguously regular from chaotic orbits, because 
some chaotic orbits may be extremely sticky and conserve their shape and frequencies 
over many orbital periods.
Secondly, such `fully mixed' models, in which all chaotic orbits with the same energy are averaged, 
cannot be built self-consistently \citep{MerrittFridman96, Siopis99}: 
the resulting super-orbit is featureless and rounder than the density distribution, 
and therefore not very suitable for the model, but regular orbits alone do not have 
a sufficient variety to support the shape, especially in the outer parts of Dehnen models.

If, on the other hand, we choose to treat chaotic orbits in the same way as regular ones in 
constructing \SM, the model is no longer guaranteed to be stationary. It means that even in
a constant potential, the shape of the mass distribution of a given
chaotic orbit may slowly evolve in time, in the process of 
chaotic diffusion. This should lead to a global evolution of the model
shape, but it is not easy to predict how strong this will be and in
what sense. 

In general, we may expect that a fully chaotic orbit fills all available configuration space 
within its corresponding equipotential surface, which is rounder than the equidensity surfaces 
of the triaxial model. 
If we somehow divide all orbits into three classes -- regular, sticky and fully chaotic -- then 
in the course of time there will be exchange of orbits between the last two classes: some sticky 
orbits may eventually become unstuck, and vice versa. The outcome of this `diffusion' depends on 
the initial orbit population of the model: if initially there were few strongly chaotic orbits 
(which is to be expected in a triaxial self-consistent model, since
such models are not very suitable 
to support the shape of the density distribution), then we may expect that their fraction 
increases in time, and the overall shape becomes rounder. 
However, the timescale for this chaotic mixing may be rather long, and the change of the overall 
shape of density distribution does not necessarily happen on this timescale.
\citet{MerrittValluri96} find that an ensemble of points initially confined to a small region in 
the chaotic part of the phase space evolves to a nearly time-invariant mixed distribution in 
$\sim 100\, T_{dyn}$. However, such mixing does not necessarily
distribute orbits uniformly over all the 
accessible chaotic part of phase space -- it may well be confined to a region surrounded by 
resonant tori, which significantly slow down further diffusion \citep{ValluriMerritt00}.

\begin{figure}  
$$\includegraphics{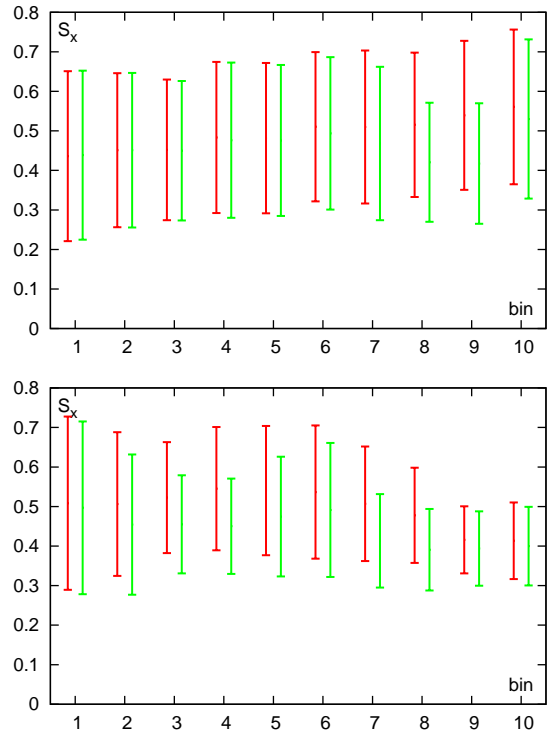}$$
\caption{
Shape change for chaotic orbits during $T=1000\,T_{dyn}$. 
Each pair of error bars shows the spread in distribution of the $x$-axis flattening $S_x$
(Eq.~\ref{eq_flattening})
for initial (left, red) and final (right, green) $100\,T_{dyn}$ intervals of time, 
averaged over the ensemble of chaotic ($\Delta\omega>10^{-3}$) orbits
in a given energy bin. The horizontal axis corresponds to 10 bins, each
of which contains 10\% of the total mass, with the innermost particles 
in the left bin.\protect\\
Top panel: $\gamma=1$, bottom: $\gamma=2$ models.
The decrease in the average value in each pair means that orbits become rounder 
(both $y$- and $z$-axis components increase at the expense of $x$-axis component). 
It is evident that in the weak-cusp case only the chaotic orbits in the outer shells do change 
shape systematically to become rounder, while in the strong-cusp case this tendency exists 
for most of the radial shells.}  \label{fig_shape_change_orbits}
\end{figure}

To study this further we perform the following experiment: we create \Nbody realizations of \SM 
with $10^4$ equal-weight particles representing a self-consistent solution, 
in which orbits were integrated for $100\,T_{dyn}$, 
Then we continue to evolve them in the same potential for 10 subsequent intervals of $100\,T_{dyn}$,
and compare the properties of each orbit in the first and the last interval.
The results show that indeed the chaotic orbits show the tendency to become rounder.
We may quantify this by measuring the change in flattening of an orbit
in each direction by using $S_i$ (eq. 4). 
Fig.~\ref{fig_shape_change_orbits} shows the mean value and spread of
flattening along the $x$ axis,
measured for the first and the last interval of $100\,T_{dyn}$, averaged over the ensemble of 
chaotic orbits (defined as those that changed their leading frequencies by more than $10^{-3}$ 
between these two intervals, as described in Sect.~\ref{sec_chaos_general}).
The systematic decrease of the $x$-flattening (and corresponding
increase in both $y$ and $z$, not shown here)
means that orbits, on average, become rounder. 
But it is important to note that this effect is not observed for all energies: 
in the weak-cusp case, only the chaotic orbits in the outer $\sim 25\%$ of energy bins were found 
to exhibit a substantial evolution, while for the strong-cusp case most chaotic orbits became 
rounder. 

\begin{figure*} 
$$\includegraphics{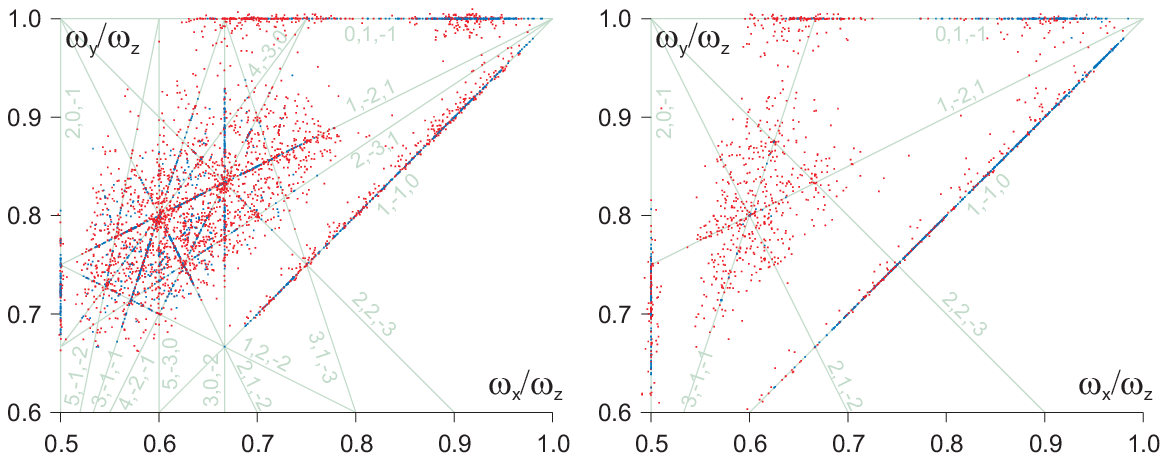}\includegraphics{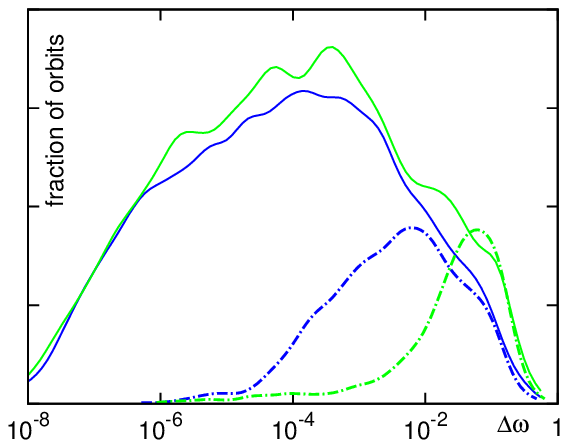} $$
\caption{ Frequency maps for the $\gamma=1$ (left) and $\gamma=2$ (middle) model.
Blue dots mark regular and red dots chaotic
orbits (as distinguished by the value of Lyapunov exponent). 
The right panel shows the FDR distribution ($\Delta\omega$) for these two
cases (blue -- $\gamma=1$, green -- $\gamma=2$), separately for
tube-like orbits (short axis tubes, long axis tubes and chaotic orbits close to $1:1$
resonances, solid lines) and for box-like orbits (dot-dashed lines). 
} \label{fig_freq_maps_y1_y2}
\end{figure*} 

The difference between these cases may be attributed to the presence of a rich network 
of resonances in the $\gamma=1$ model. 
Fig.~\ref{fig_freq_maps_y1_y2} shows frequency maps for the two models.
In the weak-cusp model there is a considerable fraction of points lying on or near resonant 
lines, corresponding to either regular or sticky chaotic orbits. Consequently, they have rather 
low values of FDR (right panel, blue), with $\Delta\omega \lesssim 10^{-2}$.
On the other hand, $\gamma=2$ model, as well as the outer parts of $\gamma=1$ model, 
have no major resonances (apart from 1:1 $x-$ and $z-$axis tubes 
and 1:2 $x-z$ banana orbits). Most non-tube orbits in these cases are quite strongly 
chaotic with $\Delta\omega \gtrsim 10^{-2}$.
We may thus infer that indeed the existence of resonances slows down
chaotic diffusion.

\begin{figure}  
$$\includegraphics{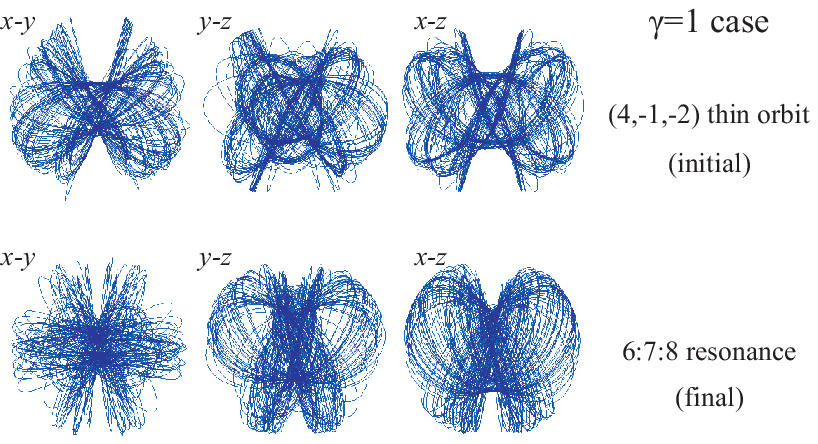}$$
$$\includegraphics{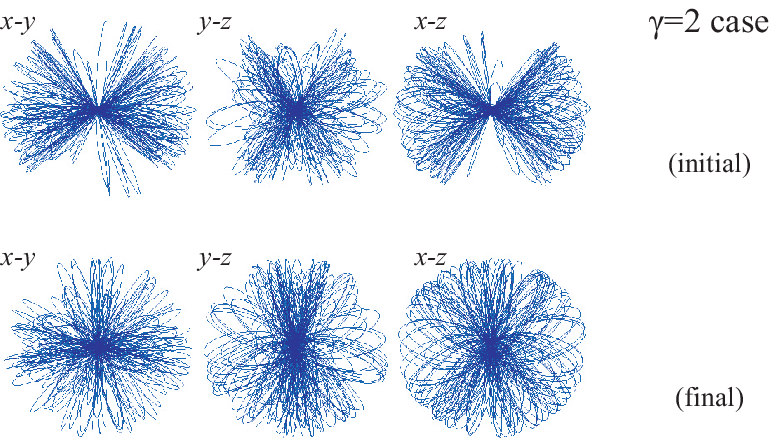}$$
\caption{
An example of shape change for two $100\,T_{dyn}$ segments (separated by $1000\,T_{dyn}$) 
of the same chaotic orbit. \protect\\
Top two rows: orbit with energy from the 2nd out of the 10 bins, in the weak-cusp model.
The initial and final orbit segments are parented by different resonances and have rather 
well-defined shapes. Even though the change in shape is substantial, there may well exist 
another orbit in the self-consistent solution which changes in the opposite sense. \protect\\
Bottom two rows: orbit with an energy from the 9th bin in the strong-cusp model. 
The initial shape is more elongated along the $x$ axis, but the final one fills almost all the equipotential 
surface, which is close to a sphere. There were no or very few such round orbits which could become 
elongated to counteract this diffusion.
}  \label{fig_shape_change_orbit_y1}
\end{figure}
 
However, it may also be considered somewhat differently,
as sketched in Fig.~\ref{fig_shape_change_orbit_y1}.
Examination of the shapes of low-energy chaotic orbits of the $\gamma=1$ model reveals that they
are mostly sticky, having a more-or-less well defined shape for any interval of $100\,T_{dyn}$. 
This means that even if an orbit is scattered to a different region of phase space, 
it may still retain a distinct shape and so represent useful `building blocks'. 
It could be expected that there is roughly a balance between orbits which 
became rounder and those which became flatter.
On the other hand, orbits from the high-energy part of the $\gamma=1$ model are 
initially closer to having a well-defined shape, but become more fuzzy-looking at the end. 
They are not balanced by the opposite flux of orbits becoming more regular, 
since there were initially very few of those near-round chaotic orbits 
in the self-consistent equilibrium.
A similar argument is used in \citet{Muzzio05} and in \citet{Aquilano07} 
to explain why it is advantageous 
to have a density model which is \textsl{less} triaxial in the outer parts, and therefore can 
adopt a large initial fraction of fully chaotic (almost round) orbits to balance this 
`puffing up' of initially more elongated chaotic orbits.
Anyway, the existence of shape-supporting resonances with their associated families of orbits, 
even chaotic, seems to be an important condition for slowing down chaotic diffusion.

On the other hand, orbits with $\Delta\omega<10^{-3}$ did not substantially change their shape 
in either case, neither individually, nor on average. 

Up to now we considered the evolution in shape during a period of a fixed number of dynamical times,  
i.e. a time which varies strongly with energy. 
We may instead focus on the changes that may occur to the model during a fixed physical time 
(say, 1000 dimensionless time units, which roughly corresponds to the
Hubble time and to the time of the \Nbody simulation which we will
discuss in Sect.~\ref{sec_nbody_evolution}).
For the weak-cusp case the dynamical time of orbits at the energies where changes in shape occur
is $T_{dyn} \gtrsim 50$; hence the chaotic diffusion occurring on timescales $\gtrsim 100\,T_{dyn}$
is essentially non-relevant on the timescale of the calculation.
However, for the strong-cusp case the orbits in the inner bins have $T_{dyn} \lesssim 1$, 
so we may expect the chaotic diffusion to have an impact on the shape evolution.

To quantify the impact that chaotic diffusion may have on the shape of a model, we used the 
following approach.
We took the same set of $10^4$ orbits from a self-consistent solution as above, 
and re-integrated them in two variants of fixed (time-independent) potential: 
a smooth Dehnen analytic potential (same as was used in \SM) 
and a frozen-\Nbody potential obtained from a rigid distribution of
$10^6$ particles representing the density of the Dehnen model.
We calculate the orbits for 10 subsequent intervals of time, 100 time units each 
(not to be confused with $100\,T_{dyn}$ in the previous experiment!).
On each of these intervals, we sample 100 points from each orbit, thereby creating 
an \Nbody model with $10^6$ particles, and record the shape of this model.
The difference from the previous experiment is that we evolved the orbits for fixed `physical' 
time, rather than for fixed number of dynamical times which depend on energy. 
This way, the inner parts were much `older' in terms of dynamical time.

The results of this calculation are presented later in Section~\ref{sec_nbody_evolution}, 
for the $\gamma=2$ case only. They show that evolution of shape during 1000 time units is substantial, 
and comparable to the evolution exhibited in the self-consistent \Nbody simulation. 
For the $\gamma=1$ case, no evolution was observed, as expected from the above consideration.

There are obvious ways to suppress this chaotic diffusion in \SM. 
First of all, one may reduce the fraction of chaotic orbits by assigning them a penalty in the 
objective function. The difficulty lies in the poorly defined distinction between regular and 
sticky orbits, and a small fraction of (weakly) chaotic orbits in the model is unavoidable.

Second, we deliberately chose a rather small interval of integration ($100\,T_{dyn}$) for \SM, 
which is not enough to ensure that chaotic orbits uniformly fill their allowed region of phase space. 
Indeed, a method proposed by \citet{Pfenniger84} consists in an adaptive selection of the integration 
time for orbits that seem to be chaotic, to improve the coverage of the available phase space: 
if the cell occupation numbers differ too much for the first and the second halves of the orbit, 
we continue the integration. The convergence, however, is not guaranteed, and can be rather slow, 
so one should impose an upper limit for integration time. 
\citet{Capuzzo07} found that even setting the maximum time equal to $2\,T_{Hubble}$ and then 
continuing the integration of orbits until $5\,T_{Hubble}$, some change of the overall shape of 
density distribution is observed. 

Third, part of the problem with chaotic orbits lies in their stickiness, which may be reduced 
at the stage of orbit integration by adding a weak noise (random force) to the equations of motion.
\citet{KandrupPS00} and \citet{SiopisKandrup00} demonstrated that even a weak noise may dramatically increase 
the rate of chaotic diffusion.

These efforts, however, are pretty much useless when we are trying to suppress the chaotic diffusion 
in an \Nbody model, since we found it very difficult, if at all possible, to preserve the 
attribute of chaoticity in transferring orbits to a slightly different potential.

\section{$N$-body evolution of models}  \label{sec_nbody_evolution}

In order to confirm our expectations about the chaotic diffusion, and to test the overall stability 
of triaxial Dehnen models, we convert our Schwarzschild model (\SM) to
an \Nbody model (\NM) by sampling  points
randomly from trajectories, their number being proportional to the weight of
the orbit in the \SM.
Our standard number of bodies in \NM is $10^6$, which means that, on average, an orbit from \SM 
with nonzero weight is sampled with $\sim 10^2$ points.

We use the \Nbody code \texttt{gyrfalcON} \citep{Dehnen00, Dehnen02} to evolve \NM for $T=1000$ \Nbody time 
units, corresponding roughly to a Hubble time ($10^{10}$ yr) for a model scaled to 
$M=3\cdot10^{11}\,M_\odot$, $a=5$~kpc.
(The half-mass (long-axis) radius of the $\gamma=1$ model is $2.4$ and the dynamical time for this 
radius is $23$ time units; for $\gamma=2$ these values are $1$ and
$6.1$, respectively).
The smoothing length was set to $\epsilon=0.01$.

First we consider only one variant of the \SM, namely with both the
fraction of chaotic orbits and the velocity anisotropy being unconstrained.
Other possibilities will be considered in section~\ref{sec_schw_variants}.

All models were initially in virial equilibrium (virial ratio $2T/W=1$
within a fraction of percent 
accuracy), and the kinetic and potential energy remained the same in the course of evolution to 
within few~$\times 10^{-3} (10^{-2})$ in the weak (strong) cusp
case. To check the stability of our model, we measured the change in
time of the following parameters:  
\begin{itemize}
\item mean-square change of energies of particles, to quantify the 
effect of two-body relaxation;
\item density profile (spherically averaged);
\item axial ratios of equidensity surfaces depending on radius;
\end{itemize}

\begin{figure}  
$$\includegraphics{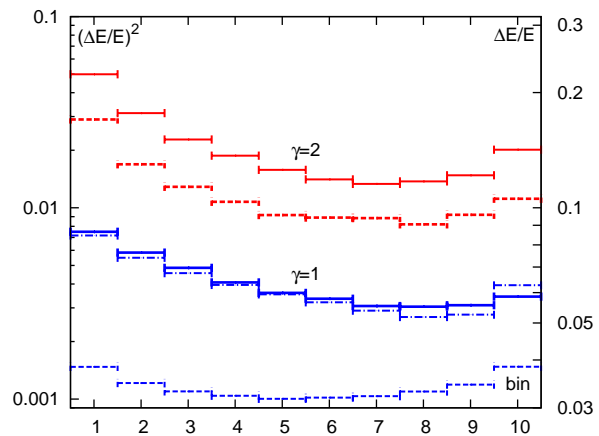}$$
\caption{Relative energy change, $\Delta E/E$, after $T=1000$, for particles in 10 energy bins
of \Nbody models (leftmost are the most bound particles). 
The red lines correspond to $\gamma=2$ models: the solid for $N=10^6$ and
the dashed for $2\cdot 10^6$. The solid and dashed blue lines
correspond to a $\gamma=1$ model, with $N=10^6$ and $5\cdot
10^6$, respectively. The blue dot-dashed line corresponds to a
$N=10^6$ model built by the iterative method (Sec.~\ref{sec_compare_iterative}).
}\label{fig_energy_diffusion}
\end{figure}

\begin{figure}  
$$\includegraphics{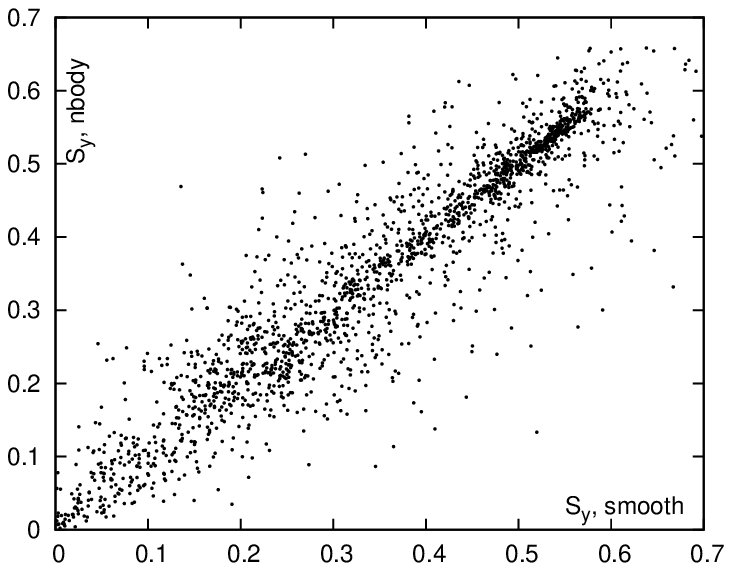}$$
$$\includegraphics{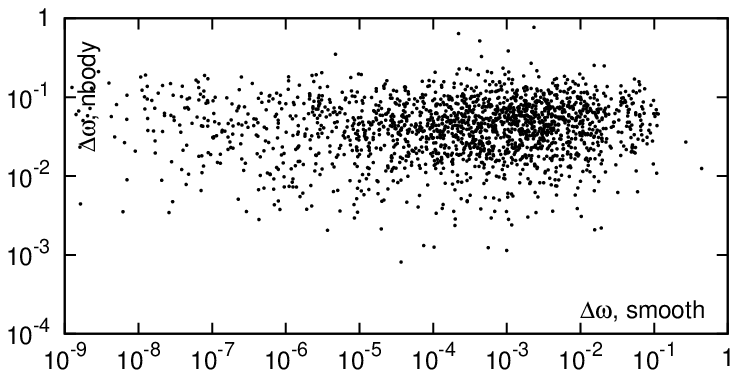}$$
\caption{
Correspondence between orbits in \SM and in \NM, for a sample of 2000 orbits in the inner 40\% of 
the weak-cusp model, which had completed at least 50 periods during the run.
\protect\\
Top panel: correlation between orbit shapes (more specifically, flattening along the $y$ axis);
the shapes of the orbits in the \Nbody run are reasonably similar to
those of the parent \SM orbits. \protect\\
Bottom panel: correlation between FDR -- meant to measure the degree
of chaos -- for the two sets of orbits. Note the difference in scale
between the ordinate and the abscissa.
In \NM the change of orbit frequencies is caused mostly by fluctuations in energy 
(see Fig.~\ref{fig_histogr_fdr_de}), and so has essentially no correlation 
with either the true degree of chaos, or the $\Delta\omega$ of the parent orbit in \SM. 
Almost all orbits in \NM have $\Delta\omega>10^{-3}$, i.e. above the threshold used for 
separating regular from chaotic orbits in the smooth potential.
}\label{fig_freq_correspondence_nbody}
\end{figure}

\begin{figure}  
$$\includegraphics{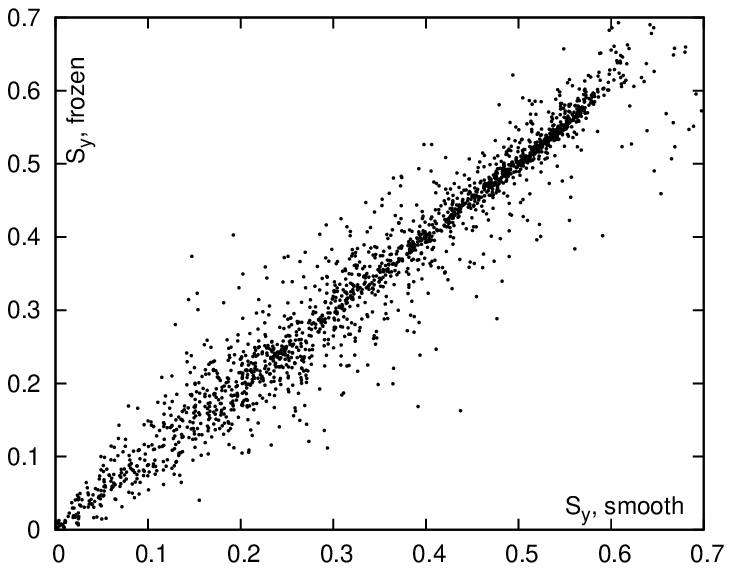}$$
$$\includegraphics{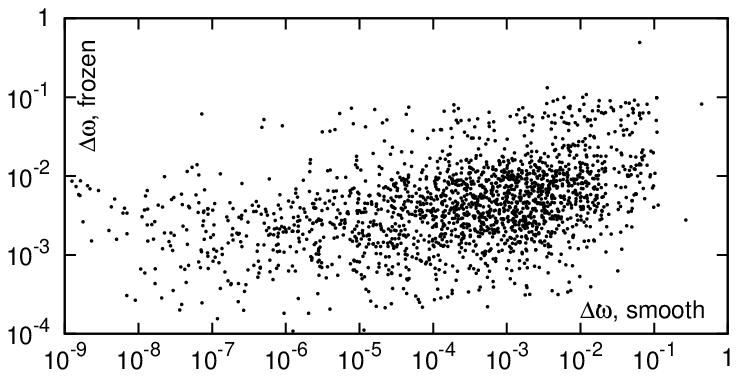}$$
\caption{
Correspondence between orbits in \SM (smooth) and frozen-\Nbody potential, 
for the same sample as in Fig.~\ref{fig_freq_correspondence_nbody}. \protect\\
The top panel displays a correlation between orbit shapes, and shows that orbits in 
the frozen-\Nbody potential resemble their counterparts in the smooth potential quite well. \protect\\
The bottom panel displays a similar correlation, but now between the FDR.
Orbits that were regular in the smooth potential (i.e. had $\Delta\omega < 10^{-3}$), 
have systematically a lower FDR in the frozen potential, although this rarely gets below
$10^{-3}$. This is due to the graininess of the potential and not to an energy error, 
which is still an order of magnitude lower.
}\label{fig_freq_correspondence_frozen}
\end{figure}

\subsection{Energy change of individual particles}

First we focus on the energy change of individual particles, which can
be due to two effects. One is the non-stationarity of the gross potential. 
This, in turn, may be either due to the model being initially not in perfect equilibrium, or due to 
large-scale dynamical instabilities. Although the Schwarzschild modelling technique is aimed at 
constructing models in equilibrium, the actual accuracy of this equilibrium may vary. 
In particular, we found that the standard practice of assigning initial conditions for orbits from 
a grid of discrete energy levels results in quite large initial fluctuations of energies.
We therefore adopted a smooth energy distribution for orbits in the \SM. 
Nevertheless, the change of energy that is due to the fact that the
model is not in perfect equilibrium initially should be confined to the initial times 
of the simulation and thus should not influence the following discussion.

The second reason for an energy change is the unavoidable two-body relaxation. It leads to
a random-walk in energy space, 
with a mean squared deviation of energy $\Delta E^2$ growing linearly with time. 
To estimate the importance of this effect, we measured the squared relative change in energy,
$(\Delta E/E)^2$, accumulated by the end of the integration time and averaged over particles 
in several energy bins 
(we checked that it indeed grew linearly with time, as expected for diffusion, apart from 
a possible initial period of faster growth if the model was not in perfect equilibrium).
Figure~\ref{fig_energy_diffusion} shows that this change remains quite
low for the weak-cusp model, 
of order $|\Delta E/E| \sim $ few percent, weakly depending on the initial energy. 
In the strong-cusp case, the relaxation rate is $\sim 5$ times stronger, although the simulation time 
is still much shorter than the relaxation time for all energies.

\begin{figure}
$$\includegraphics{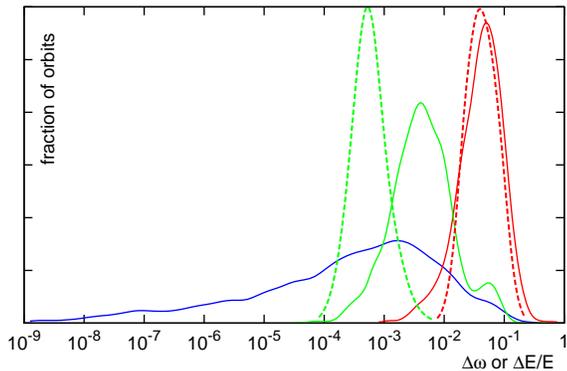}$$
\caption{
Histograms of FDR ($\Delta\omega$, solid lines) and of energy conservation error 
($\Delta E/E$, dashed lines) for orbits from Figs.~\ref{fig_freq_correspondence_nbody} 
and \ref{fig_freq_correspondence_frozen}.
The blue line corresponds to orbits in the smooth potential, green one
to orbits in the frozen-\Nbody potential, and the red line to orbits
in the live simulation.\protect\\
For the live simulation the energy conservation error is quite large, between $10^{-2}$ and $10^{-1}$, 
and it determines the error in the frequency estimation. 
For the frozen-\Nbody potential the energy conservation is still far from being perfect 
(of order $\sim 10^{-4}..10^{-3}$, due to the approximations of the tree-code
and the finite number of particles), but good enough not to be the major 
source of frequency diffusion. 
}\label{fig_histogr_fdr_de}
\end{figure}

\begin{figure}  
$$\includegraphics{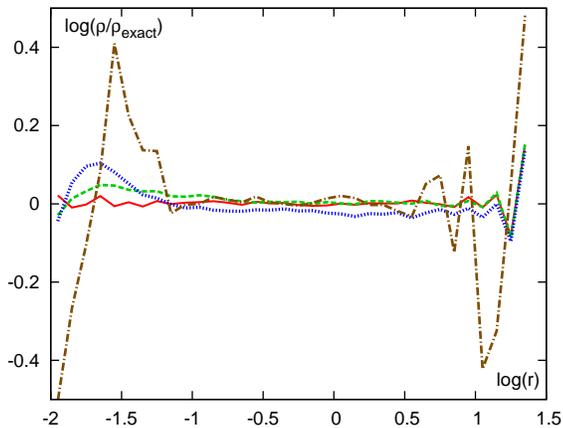}$$
\caption{Density profile evolution for the $\gamma=2$ model. We  
plot the difference between density in the model and the exact value
($\log(\rho/\rho_{exact})$) as a function of radius ($\log(r)$).
The red solid curve corresponds to the initial model, which is fairly
close to the exact Dehnen profile and
the green dashed curve corresponds to the model after it evolved over
10 time units. From the latter we see that the smoothing reduces the density at $r \lesssim \epsilon$ 
and distributes the excess of mass slightly further from centre.  
The blue dotted line corresponds to the end of simulation, where the axial ratios have changed substantially.
For comparison, the brown dot-dashed line gives the initial density in \SM with initial 
conditions assigned on a regular grid at 20 fixed energy levels (see
Sec.~\ref{sec_schw_variants}) and clearly 
deviates substantially from the exact profile at small and large radii.
}\label{fig_density_evol}
\end{figure}

This, however, leads to an important implication for the detection of chaos by the frequency 
diffusion rate. In fact, for a perfectly regular orbit $\Delta\omega \simeq \Delta E/E$, and so 
the lower bound on FDR depends on the energy conservation of a given
orbit and the numbers shown in Fig.~\ref{fig_energy_diffusion}
show that it is indeed quite high. In the initial distribution of FDR 
(in the smooth potential of \SM) there were no more  
than a few percent of orbits with FDR $\Delta\omega > 0.03$, while in
the \NM they comprise the majority of orbits. 
It therefore does not make sense to use $\Delta\omega$ as an indicator of chaos for orbits in 
a live \Nbody simulation. To achieve an acceptably low energy diffusion rate of 
$\Delta E/E \simeq 10^{-3}$, we would need to have of the order of
$10^{9}$ or $10^{10}$ particles
(based on the rates for squared energy change from Fig.~\ref{fig_energy_diffusion} 
being inversely proportional to the number of particles).

Not surprisingly, there is essentially no correlation between the FDR
in the smooth-potential and in 
the \Nbody models (Fig.~\ref{fig_freq_correspondence_nbody}). 
This casts a shadow on the usefulness of attempts to control the amount of chaos in the \NM 
by changing the properties of the \SM. The tests in the next section
confirm that indeed the evolution of the \NM is basically independent
of whether most orbits in \SM are regular or irregular. 
On the other hand, orbit shapes in the \NM are mostly close to those
of their parent 
orbits from the \SM, as can be seen from the top panel of
Fig.~\ref{fig_freq_correspondence_nbody}. Furthermore, we
found no correlation between the change of orbit shape in the \NM 
(which may be taken as an indicator of its chaotic behaviour) and its 
$\Delta\omega$ either in the \NM, or in the \SM (not shown here). 

It is interesting to compare the orbits in the live \Nbody models with the corresponding 
orbits in the frozen \Nbody potential (Fig.~\ref{fig_freq_correspondence_frozen}). 
The latter share with the former the energy conservation error caused by the tree-code 
force approximation and by finite-difference integration errors, but this can be kept 
as low as necessary by choosing a suitable tree cell opening angle $\theta$ and 
integration timestep. 
For our runs this energy error is between $10^{-4}$ and $10^{-3}$ for
all orbits, i.e. 
much lower than the energy diffusion in the live simulation which, as
shown in Fig.~\ref{fig_histogr_fdr_de}, is of the order of $10^{-2}$
or $10^{-1}$. 
Yet $\Delta\omega$ is an order of magnitude larger than the energy error 
and we checked that it does not change noticeably when we increased
energy conservation accuracy. This shows that
the main contribution to $\Delta\omega$ comes from graininess of the potential, 
and for most orbits is not caused by the non-integrability of corresponding smooth potential.
Note in particular that for the case of a spherically symmetric frozen-\Nbody potential, $\Delta\omega$ is also 
between $10^{-3}$ and $10^{-2}$ for the majority of orbits, which confirms that this lower limit 
is caused by graininess, not by `real' chaotic properties of the potential).
Similar values of $\Delta\omega$ were found in \citet{Valluri10} for orbits in spherical 
NFW potential represented by $10^6$ frozen particles (their fig.~1).

\subsection{Evolution of global quantities}

\begin{figure}  
$$\includegraphics{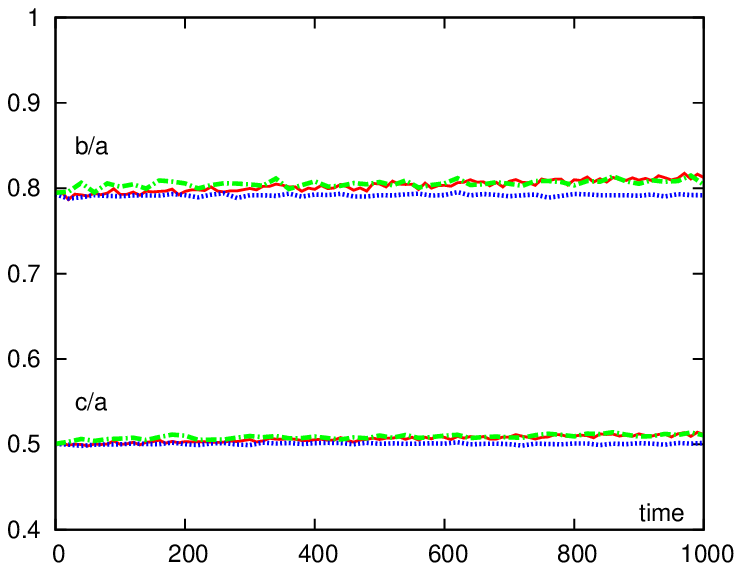}$$
$$\includegraphics{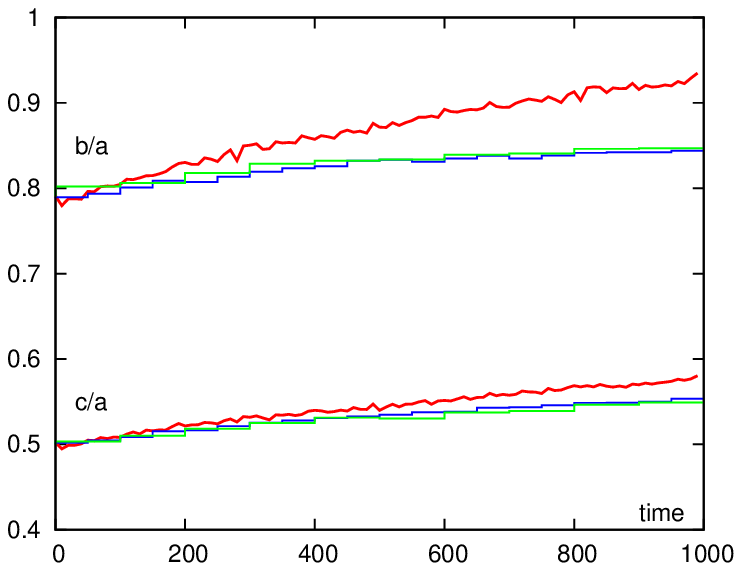}$$
\caption{
Evolution of the $b/a$ and $c/a$ axial ratios, starting from 0.79 and 0.5, respectively.
\protect\\
The top panel compares two $\gamma=1$ models. The red 
solid line corresponds to a $N=10^6$ reference run, the dotted
blue line to a $N=5\cdot 10^6$ 
high-resolution run and the green dash-dotted line to a $N=10^6$ model built
with the iterative method. 
They all show little evolution, and the high-resolution run conserves
its shape almost perfectly.
\protect\\
The bottom panel shows that the $\gamma=2$, $N_p=10^6$ simulation (red solid
line) shows a substantial evolution of shape.
It is compared to the evolutions of orbits in the fixed potential
(blue -- smooth Dehnen, green -- frozen \Nbody potential),
which demonstrate similar amount of shape change exclusively due to
chaotic diffusion.
}\label{fig_axis_ratio_evol}
\end{figure}

For all models considered, the velocity distribution (not shown) and the density profile
(Fig.~\ref{fig_density_evol}) do not show substantial evolution,
except for an unavoidable smoothing of the cusp at $r\lesssim \epsilon$.
The most important changes occur in the axial ratios of the model (Fig.~\ref{fig_axis_ratio_evol}).

There are two different methods to measure the axial ratio and its dependence on radius, 
and they give very similar results in our case.
The first method \citep{AthanassoulaMisiriotis02} is to sort particles in density and bin them 
into $N_b$ bins, then calculate the principal axes of the inertia tensor of all particles belonging 
to a given bin (if the density is a smooth monotonic function of radius, these bins are 
roughly ellipsoidal shells).
The second method -- a variation of that used by
\citet{DubinskiCarlberg91} -- 
consists in binning particles into layers bounded by concentric ellipsoids
whose axes are determined by the following iterative procedure\footnote{
A more extended discussion is found in \citet{Zemp11}.}. Consider first the inner shell 
containing $1/N_b$ of the total mass. We search for an ellipsoid with axes $a_0, b_0, c_0$ such 
that the moment of inertia of all particles within this ellipsoid has the same axis ratio as this 
bounding ellipsoid. We start from a spherical shell and find the moment of inertia of particles 
within this radius, then change the axial ratio of the bounding ellipsoid to these values and 
repeat the iteration until convergence is achieved. We then remove the inner particles from 
consideration and repeat the procedure for the next shell. In practice, the number of shells 
should be rather small to avoid `shell-crossing' and the divergence of iterations.

In what follows, we usually present the axis ratios for the 2nd out of 4 bins 
(particles between 25 and 50\%), which contains particles close enough to the centre yet 
not much affected by softening. (Indeed, for the $\gamma=2$ model the axial ratio changed 
with almost the same rate at all radii).

\begin{figure}  
$$\includegraphics{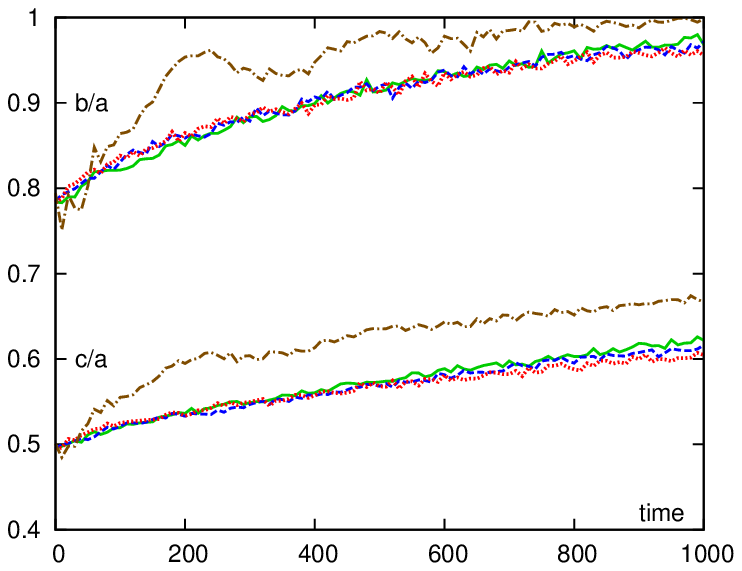}$$
$$\includegraphics{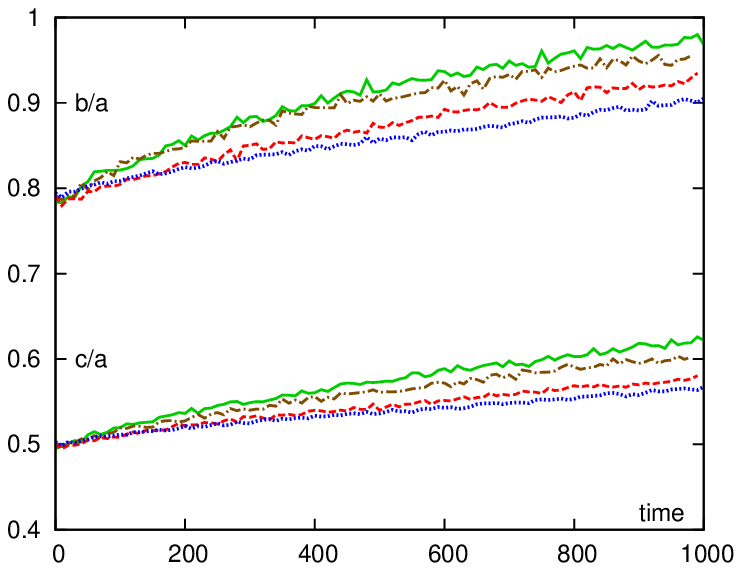}$$
\caption{
Evolution of the $b/a$ and $c/a$ axial ratios, starting from 0.79 and
0.5, respectively, for variants of the $\gamma=2$ model. 
\protect\\
The top panel shows $N_p=5\cdot 10^5$ particles \NM created from 
\SM having $N_s=20$ shells and $N_o=10^4$ orbits integrated for 100 orbital periods. 
The green solid lines correspond to the unconstrained model, 
the blue dashed line to the model with a preference of regular orbits, 
the red dotted line to the model with a preference of chaotic orbits, 
and the brown dot-dashed to the model created from grid initial
conditions (the other three models have random IC). 
This last one clearly displays the strongest shape evolution, while there is 
almost no difference between `mostly regular', `mostly chaotic' and 'unconstrained' models.
\protect\\
The bottom panel shows the effect of particle number in \NM and of number of orbits and shells in \SM.
The green solid line shows the shape evolution of a model with $N_s=20, N_o=2500$ 
and $N_p=5\cdot 10^5$ integrated for 100 dynamical times (same as in the top panel). 
The brown dot-dashed shows the $N_s=50, N_o=15000$ and $N_p=5\cdot 10^5$ model
integrated for 100 dynamical times. 
The red dashed line corresponds to the $N_s=50, N_o=12500$ and $N_p=10^6$,
integrated for 500 dynamical times 
(same as the red solid curve in the bottom panel of Fig.~\ref{fig_axis_ratio_evol}).
The blue dotted line corresponds to a model similar to that of the red line, 
but with $N_p=2\cdot 10^6$. \protect\\
This figure demonstrates that increasing the number of particles does slow down the shape evolution, 
and that at fixed $N_p$ a \SM with more shells and orbits behaves better.
}\label{fig_axis_ratio_evol_y2}
\end{figure}

For the weak-cusp model (Fig.~\ref{fig_axis_ratio_evol}, top panel) the change of axial ratio 
is very slight for the `standard' model, and virtually zero for high-resolution run with 
$5\cdot10^6$ particles, which confirms our expectations from the previous section.

In the strong-cusp case (bottom panel), however, the changes are much more obvious. 
In part, they may be attributed to the much more efficient two-body relaxation in the centre. 
However, chaotic diffusion may also play a substantial role in this, as demonstrated by 
integrations in the corresponding fixed potential (Section~\ref{sec_chaotic_shape_change}).
As seen from the lower panel of Fig.~\ref{fig_axis_ratio_evol}, the rate of change of 
the axial ratio in the fixed-potential integration is comparable to, or, more precisely,
about half that of the live \Nbody model.

\section{Exploring the variants of Schwarzschild models}  \label{sec_schw_variants}

As is well known, Schwarzschild modelling is a rather flexible approach, in the sense that 
there are many ways to construct models satisfying a given density profile. 
In fact, the non-uniqueness of the solution may sometimes be an unwanted property of the \SM, 
since there may be no a priori way to tell which one is preferred.
Here we explore how the variation of model parameters affects its `quality' and stability.
In this section we consider only the $\gamma=2$ case, which displays a
more rapid evolution.

The first parameter to vary is the velocity anisotropy. This need not be constant with radius, 
but the most obvious effect occurs when we change its value at the centre.
For $\gamma=1$, the radial-orbit instability is triggered when $\beta$ rises to 0.4, with the axial 
ratios dropping quickly (withing few time units) from 0.8 to 0.6
($b/a$) and from 0.5 to 0.4 ($c/a$). 
This confirms the results of \citet{Antonini09}. 
For $\gamma=2$, the instability occurs promptly for $\beta=0.5$, 
with the axis ratios dropping to $b/a=0.5$ and $c/a=0.33$, but it happens, 
albeit with less dramatic results, even for $\beta$ as small as 0.1, 
when the ratios instantly drop by a few percent. 
The subsequent evolution of axis ratios is slow and similar to the case with no short-term instability.

Next, we fix the velocity anisotropy profile to a $\beta$ growing linearly with the shell number 
from 0 in the centre to 0.6 in the outer parts, so that the model
becomes robust against the radial orbit  
instability, and study the effect of changing the relative fraction of chaotic orbits in the \SM.
There is still considerable freedom in distributing orbit weights between regular and chaotic orbits.
By inserting penalty terms to the objective function that gives preference to regular or to chaotic 
orbits, a solution may be constructed having a fraction of regular orbits anywhere between 
30(35)\% and 90(75)\% for the $\gamma=1(2)$ cases, respectively, with the `unconstrained' solution 
yielding $\sim 60\%$ of regular orbits. (These numbers are given for 
the \SM integrated for 100 dynamical times; for $T=500$ models the available range is narrower).
The available interval for the fraction of regular orbits is narrower for the strong-cusp case
because of the scarcity of stable orbit families in this potential. 
Most importantly, no solution having only regular orbits may be constructed in both cases.

Other factors to explore include the number of shells $N_s$ and of orbits $N_o$ in \SM, 
and the `coverage' of the accessible phase space by orbits. 
The latter factor measures how well and how uniformly an orbit samples its available phase space 
(invariant 3-dimensional torus for a regular orbit, or a higher-dimension volume 
of phase space for a chaotic one). 
As the chaotic orbits sometimes exhibit strong stickiness, it has been proposed to add some noise 
to the equations of motion to enhance the `diffusion rate' \citep{KandrupPS00}, or -- alternatively 
-- to use a `dithering method', proposed in \cite{vdBosch08}: integrating a bunch of adjacent orbits 
with slightly perturbed initial conditions and use averaged values of cell occupation times. 

In addition, we consider the effects of varying the parameters of \NM: 
number of particles $N_p$ and smoothing length $\epsilon$.

The conclusions from these studies are the following. We concentrate mainly on the axis ratio 
evolution rate (Fig.~\ref{fig_axis_ratio_evol_y2}), as it is the most apparent indicator of model 
instability. 

First, all attempts to control the evolution by changing the amount of chaotic orbits in 
the \SM fail miserably -- all three models (preferentially regular, chaotic or unconstrained) 
evolve at the same rate. 
Secondly, models with more nonzero weight orbits in the \SM  
and/or with more radial shells, tend to perform better, presumably because they
are closer to equilibrium and because their distribution function is smoother in phase space.
A particularly striking demonstration of the necessity of such smoothing is the very poor 
behaviour of the model created with the traditional method of assigning initial conditions 
on a regular grid of points in two start spaces (stationary and principal-plane) 
at a small number of fixed energy levels.

Moreover, improving the coverage of the phase space available for individual orbits by adding noise, or
increasing integration time, or averaging the contribution of a bunch of nearby orbits, 
also does not seem to help.
This conclusion may seem to be in contradiction with the results of \citet{SiopisKandrup00, 
KandrupSiopis03}, which indicate that adding noise does increase the chaotic diffusion at the 
stage of \SM, which should in principle lead to a more stable \NM. 

The reason behind this apparent contradiction is probably the following. 
These methods indeed may improve coverage of phase space for some chaotic orbits, 
which then become closer to a `fully mixed' ergodic orbit.
But since the solution cannot be obtained using only regular and well-mixed chaotic orbits%
\footnote{This, however, may well be a limitation imposed by our choice of density profile: 
\citet{Terzic02} had no trouble constructing scale-free cusps using only regular orbits, 
and \citet{Muzzio05} argue that a model with varying axial ratios (rounder in the outer parts) 
may be better adapted to having a necessary `equilibrium' population of strongly chaotic orbits.},
we still need some chaotic orbits which retain more or less distinct shape 
and do not fill uniformly their available phase space 
(in some cases, we need to increase the total number of orbits in \SM to get a 
feasible solution).
But nothing prevents such orbits from experiencing the same type of chaotic 
diffusion in the perturbed \NM potential, and since they comprise in total 
approximately the same fraction of orbits, regardless of the details of 
construction of \SM, the resulting evolution of \NM will be more or less the same. 
This argument also applies in cases where we vary the fraction of
chaotic orbits in the \SM, because this 
formal selection criterion will also consider as regular those orbits that did not 
seem to be sufficiently chaotic, but may readily become so in the \NM.
It is important to note that all these conclusions are valid also 
for integration in the fixed potential.

As for the \NM parameters, increasing either the number of particles $N_p$, 
or the smoothing length $\epsilon$ does slow down the shape evolution. 
The latter effect may be attributed to the fact that larger smoothing 
destroys the strong cusp that is responsible for chaotic diffusion.
Therefore, the dynamical properties and phase space structure of such a 
smoothed model are different from the ones in \SM.
The increase of particle number does slow down two-body relaxation, 
but this can not be the main reason for the slow-down of the shape
evolution because the timescale for chaotic diffusion is in any case
much shorter  
than the relaxation time (as seen from Fig.~\ref{fig_energy_diffusion}, 
the latter is $\sim 20-50$ times longer than the integration time 
even for $N_p=10^6$).
Instead, it must be the decrease of the potential graininess due to
the increase of the particle number that will be responsible for the
increase of the chaotic diffusion time-scale.
However, there is probably a fundamental lower limit of the rate of chaotic diffusion, 
established by the fixed-potential integration experiment (which excluded two-body relaxation), 
and indeed there seems to be much less difference between $10^6$ and $2\cdot 10^6$ runs than between 
$5\cdot 10^5$ and $10^6$. 

\section{Comparison with the iterative method for constructing equilibrium models}  \label{sec_compare_iterative}

The iterative method for constructing equilibrium models \citep{Rodionov09}
is designed to create an \Nbody model in a stable equilibrium, satisfying given constraints, 
such as specific density profile, shape, and/or kinematical constraints. 
This is achieved by performing a series of short-time integrations of
any initial \NM, and adjusting its properties after each iteration to satisfy the constraints. 
If this process converges to a solution, it will be stationary in the short term,
being in dynamic equilibrium and not having fast growing instabilities. 
Thus it is worthwhile to compare the properties and evolution of models created with 
these two different methods -- the iterative and the Schwarzschild. 
We consider here the weak-cusp ($\gamma=1$) case, 
as we were unable to create a sufficiently stable model for $\gamma=2$ using the iterative method.

The model built with the iterative method has $10^6$ particles and a
velocity anisotropy ranging from 0 in the centre to 0.7 in the outer parts, 
similar to our Schwarzschild model.
As seen from Fig.~\ref{fig_energy_diffusion}, these two models have very similar energy diffusion 
rates, which confirms that \SM is indeed in good dynamic 
equilibrium (since the model built with iterative method should be in equilibrium by definition).
The rate of shape evolution is approximately the same for these two types of models 
(Fig.~\ref{fig_axis_ratio_evol}), and the orbit population of the
model created with the iterative method is similar to that of \SM,
being roughly 60\%/10\%/30\% for short-, long-axis tubes and other orbits, correspondingly.

Therefore, we may conclude that the iterative and Schwarzschild methods give similar results, 
despite being conceptually very different. This supports the idea that orbital content and other 
properties of a model do not depend substantially on the way it was
constructed, provided this was adequate, but depend mainly on 
the intrinsic properties of the potential.

\section{Discussion and conclusions}  \label{sec_conclusion}

We studied in detail two triaxial Dehnen models, one with cusp slope $\gamma=1$ (weak cusp) 
and the other with $\gamma=2$ (strong cusp), both constructed with the Schwarzschild method.
Our goal was to check the stability of these models by direct \Nbody simulations
(with a number of particles $N_p\ge 10^6$), and to explore how the chaotic orbits influence 
the long-term evolution of model shapes.

The $\gamma=1$ model demonstrated a remarkable stability over the simulation timescale 
($\sim 50$ half-mass dynamical times), which confirms earlier results of \citet{Holley01} 
obtained, however, for a less triaxial model and for a shorter evolution time,  
and these of \citet{Muzzio09} for a cuspy ($\gamma\simeq 1$) model
with a similar triaxiality and a quite large ($\ge 50\%$) fraction of chaotic orbits. 

The $\gamma=2$ model, on the contrary, displayed substantial shape evolution, 
increasing the axis ratio $b/a$ from 0.8 to $>0.9$ and $c/a$ from 0.5 to $>0.55$ 
in a simulation time corresponding to $\sim 200$ half-mass crossing times. 
We attribute this evolution to chaotic diffusion of orbits, which
leads to a similar rate of shape 
change even for integration in a fixed potential (smooth or frozen-\Nbody). 

The difference between these two models may be explained by the different phase space structure of 
the underlying potential. In the weak-cusp case there are numerous resonant and thin orbit families, 
and most chaotic orbits are in fact quite sticky. A similar conclusion about the importance of 
resonances in preventing chaotic diffusion from changing significantly the overall shape of the 
model was reached by \citet{Valluri10}.
On the contrary, in the strong-cusp case the phase space is relatively `simple', with more 
strongly chaotic orbits which are rounder in shape. 
This is in good agreement with the results of \citet{MerrittQuinlan98}, \citet{Holley02} 
and \citet{Kalapotharakos04} who find that the triaxiality is rapidly 
destroyed in the presence of a large enough central point mass, which has analogous effect to 
a strong cusp in terms of degree of chaos \citep{ValluriMerritt98}. 

However, the very definition of a chaotic orbit is quite problematic
for an \Nbody model: the only 
available measure, the frequency diffusion rate ($\delta\omega$), is
mainly determined by the change of particle energy during the simulation. 
Almost all particles in the simulation have $\delta\omega>10^{-2}$, and this quantity has no 
correlation with the $\delta\omega$ for orbits with the same initial
conditions in the smooth potential, 
even though the shape of an orbit is very similar in the two cases. 
By contrast, in a frozen-\Nbody potential $\delta\omega$ is typically in the range 
$10^{-3}$ to $10^{-2}$ for orbits which are regular in the smooth potential, and is caused
by graininess of the potential. 

An important conclusion that we reached is that constraining the proportion 
of stars on chaotic or regular orbits in the \Nbody model is not of
much use for increasing the model stability. 
The first reason comes directly from the impossibility to detect chaos for orbits in the \Nbody 
simulation, discussed in the previous paragraph.
The second reason is that there is not much freedom in getting fundamentally different orbit 
population in \SM, when we constrain both the density and velocity anisotropy 
profile and try to change the amount of stars on chaotic orbits.
All that happens really is that we may privilege, in the orbit selection, some inherently chaotic 
orbits which did not demonstrate \textsl{sufficiently chaotic} behaviour during the orbit integration 
over those that did; 
however, in the \Nbody model both kinds of orbits will be slightly different from their 
counterparts in \SM, and will probably have almost equal chance 
to contribute to chaotic diffusion.
Therefore, \textsl{it makes no sense to reduce the fraction of chaotic orbits in \SM} in an attempt 
to improve the quality (stability) of \NM; this is determined
mainly by the orbital structure of the 
underlying potential, and to some degree by a careful construction of \SM.
We found that increasing the number of orbits in \SM does help to reduce evolution, and that 
randomly assigned initial conditions in \SM produce a better model than the conventional method of 
drawing them from regular grid of points on fixed energy levels.

The arguments about the impossibility of distinguishing between regular and (weakly) chaotic orbits 
and controlling the nature of any particular orbit in \Nbody simulation may, at first sight, 
seem to be grounded on the limited resolution of the simulation and inaccuracies introduced by 
energy relaxation.
However, while the two-body relaxation times in real galaxies are many orders of magnitude 
longer than in our runs, there are always some large-scale processes that both shuffle stars across 
phase space and change properties of this phase space, at least to
some degree. As such we can mention encounters of stars with globular
clusters, or with molecular clouds or with transient spiral
segments, which will perturb their trajectories as much as, if not more
than they would be by the lower number of bodies in our idealized \Nbody simulations. 
Therefore, it is unrealistic to expect that a carefully arranged initial composition of stars 
mostly on regular orbits will be preserved over a Hubble time. 
Rather, an unconstrained (in terms of fraction of chaotic orbits) arrangement of orbits, 
evolving according to the gross properties of the underlying potential, 
should give a better idea of the typical rate of shape change.

The difference in the evolution of weak- and strong-cusp models -- the latter 
becoming noticeably rounder during the Hubble time -- is also in agreement 
with observations that show that fainter elliptical galaxies which are, on average, 
more cuspy, have more spherical shapes than brighter ones, which have 
shallower density profiles and are more triaxial \citep[e.g.][and
references therein]{TremblayMerritt96}.

Apart from the cases when radial-orbit instability occurs, and from
the cases displaying secular shape evolution due to chaotic diffusion, 
models built with the Schwarzschild method appear to be in stable equilibrium.

\textbf{Acknowledgments:} We are grateful to S.~Rodionov for preparing the 
model built with the iterative method, to F.~Antonini and D.~Merritt for 
fruitful discussions, and to the referee for helpful remarks.
EV was partially supported by Russian Ministry of science and education 
(grants No.2009-1.1-126-056 and P1336), and by the CNRS during his visit to 
the Laboratoire d'Astrophysique de Marseille (UMR6110).

\label{lastpage}

\end{document}